\def\fr#1#2{{#1\over #2}}

\def\pd{\partial}                

\def\a{\alpha}           \def\b{\beta}           \def\th{\theta}
\def\m{\mu}              \def\n{\nu}             \def\k{\kappa}
           \def\g{\gamma}          \def\ve{\varepsilon}
\def\D{\Delta}           \def\d{\delta}          \def\r{\rho}
\def\S{\Sigma}           \def\s{\sigma}          \def\t{\tau}
\def\O{\Omega}                     
          \def\l{\lambda}         \def\Th{\Theta}
        \def\cH{{\cal H}}

  \def\hpd{\hat\partial}

\def\bpsi{\bar\psi}
\def\psm{\psi_-}
\def\psp{\psi_+}
\def\pspm{\psi_{\pm}}
\def\cH{\cal H}

\documentstyle[abstract,aps,eqsecnum]{revtex}

        \def\nn{\nonumber}
\def\be{\begin{equation}}             \def\ee{\end{equation}}
\def\bea{\begin{eqnarray} }           \def\eea{\end{eqnarray} }
\def\ba#1{\begin{array}{#1}}          \def\ea{\end{array}}
\def\lab#1{\label{#1}}                \def\eq#1{(\ref{#1})}
\def\bsubeq{\begin{mathletters}}      \def\esubeq{\end{mathletters}}
\def\bitem{\begin{itemize}}           \def\eitem{\end{itemize}}

\begin{document}
\draft
\title{Canonical approach to 2D WZNW model, non-abelian bosonization
       and anomalies} 
\author{B. Sazdovi\'c\thanks{E-mail address: sazdovic@phy.bg.ac.yu} }
\address{Institute of Physics, 11001 Belgrade, P.O.Box 57, Yugoslavia}
\date{\today}
\maketitle
\begin{abstract}

The gauged WZNW model has been derived as an effective action,
whose Poisson bracket algebra of the constraints is isomorphic to the
commutator algebra of operators in quantized fermionic  theory.
As a consequence, the hamiltonian as well as usual lagrangian
non-abelian bosonization rules have been obtained, for the chiral
currents and for the chiral densities. The expression for the anomaly
has been obtained as a function of the Schwinger term, using canonical methods.
\end{abstract}

\pacs{PACS number(s): 04.60.Ds, 11.10.Kk, 11.15. q}

\section{Introduction} 

It is well known that in $1+1$ dimensions there exist equivalence
between fermi and bose theories in the abelian \cite{1} and
non-abelian case \cite{2}. In the later Witten
demonstrated that free field theory of $N$ massless Majorana
fermions is equivalent to the non-linear sigma model with
Wess-Zumino term at the infrared-stable fixed point, because both
theories obey  the same Kac-Moody (KM) algebras.
The extension of this equivalence have been considered by several authors
\cite{3}. They introduced external chiral gauge fields and showed the identity
of the effective actions which implies the identity of correlation
functions.

In this paper, starting with non-abelian fermionic theory coupled with
background gauge fields, we are going to construct the equivalent
bosonic theory for general gauge group. Our approach is different from the
previous one and naturally works in the {\it hamiltonian} formalism.
We believe that it gives a simpler resolution of the problem.

The classical fermionic theory is invariant under local
non-abelian gauge transformations. Consequently, the first class
constraints (FCC) $j_{\pm a}$ are present in the theory and satisfy
non-abelian algebra as a Poisson bracket (PB) algebra. In the quantum
theory the {\it central term} appears in the commutator algebra of the
operators ${\hat j}_{\pm a}$, so that the constraints become second
class (SCC) which implies the existence of the anomaly \cite{4}.
These known results will be repeated in Sec. II for completeness
of the paper and in order to fix our notation.

We define the effective bosonized theory, as a classical theory
whose PB algebra of the constraints $J_{\pm a}$ is isomorphic to the
commutator algebra of the operators ${\hat j}_{\pm a}$, in the
quantized fermionic theory. This is the way how the bosonized theory
at the classical level incorporates anomalies of the quantum
fermionic theory.

In Sec. III we find the effective action $W$, for given algebra as its
PB algebra. The similar problems has been considered before in the
literature \cite{5}. Using the method of coadjoint orbits, they
showed that KM algebra yield the Wess-Zumino-Novikov-Witten (WZNW)
model. Here we are going to present a new, canonical approach.
We introduce the phase space coordinate $q^\a , \pi_\a$ and
parameterize the constraints $J_{\pm a}$ by them. One of the
main points of the paper is to find the expressions for the
constraints $J_{\pm a}$ and for the canonical hamiltonian
${\cH}_c$ in terms of phase space coordinate, satisfying a specific
PB algebra. We then use the general canonical
method \cite{6,7} for constructing the effective action $W$
with the known representation of the constraints. By eliminating the momentum
variables on their equations of motion we obtain the bose theory in the
background fields, which is equivalent to the quantum fermi
theory in the same background. This bose theory is known as a
{\it gauged WZNW action}.

In Sec. IV we derive {\it hamiltonian } non-abelian bosonization
rules. It is easy to obtain the formulae for the  currents, just
differentiating the functional integral with respect to the
background fields. We also derive the rules for $\bpsi \psi$ and
$\bpsi \g_5 \psi$ terms, using the approach of this paper. Note that
our hamiltonian bosonization formulae for the currents depend on
the momenta, while those for mass term depend only on the
coordinates. Witten's non-abelian bosonization rules can be
obtained from the hamiltonian ones, after eliminating the momenta.

In Sec. V we obtain the expression for the anomaly, using
canonical method. We extend the general canonical
formalism, from systems with FCC to the systems with SCC
where the central term appears. We find the expressions
for the left-right, as well as for the axial anomaly.

Sec. VI is devoted to concluding remarks. The derivation of the
central term, using normal ordering prescription is presented
in the Appendix.

\section{Canonical analysis of the fermionic theory} 

\subsection{Classical theory}

Let us consider the theory of two dimensional  massless Majorana
fermions $\psi^i \, (i=1,...N)$, interacting with the external
Yang-Mills fields $A_\mu$ and $B_\mu$, with the action
\be
S= \int d^2\xi [\bpsi i \hpd \psi-i \bpsi {\hat A}{1 + \g_5 \over 2} \psi
 -i \bpsi {\hat B}{1 - \g_5 \over 2}\psi] \,  .      \lab{2.1}
\ee
We can rewrite it as
\be
S=\int d^2\xi[ i\psm^* {\dot \psm} +i\psp^* {\dot \psp} +i\psm^*
\psm' -i\psp^* \psp' -i\sqrt{2}(A_+^a \psm^* t_a \psm +B_-^a
\psp^* t_a \psp)] \,    .                                  \lab{2.2}
\ee
We chose anti-hermitian matrices $t_a$ as the generators of the gauge
group $G$, introduce light-cone components $V_\pm ={1 \over \sqrt{2} }
(V_0 \pm V_1)$ for the vectors, and write the gauge potentials as $A_+ =A_+^a t_a$
and $B_- =B_-^a t_a$. We use the basis $\g^0=\s_1 \, , \g^1=-i \s_2 \,
,\g_5 =\g^0 \g^1 =\s_3$, and define the Weyl spinors by  the conditions
$\g_5 \pspm =\mp \pspm $. For simplicity, we write $\pspm^* \pspm$ and
$\pspm^* t_a \pspm$ instead of $\sum_i \pspm^{* i} \pspm^i$ and
$\sum_{ij} \pspm^{*i} t_{aij} \pspm^j$.

The fermionic  action \eq{2.2} is already in the hamiltonian form and
we can conclude that there are two basic lagrangian variables $\psm^i$ and
$\psp^i$ appearing with time derivative, whose conjugate momenta are
$\pi_{\pm}^i=i\pspm^{* i}$. Variables without time derivatives, $A_+^a$ and
$B_-^a$, are lagrange multipliers and the primary constraints
corresponding to them are the currents
\be
j_{\pm a}=i\pspm^* t_a \pspm = \pi_\pm t_a \pspm \,  .        \lab{2.3}
\ee
The canonical hamiltonian  density can be expressed in terms of
the chiral quantities $\th_\pm$
\be
{\cH}_c = \th_+ -\th_-  \qquad  ( \th_\pm= i \pspm^* \pspm'=\pi_\pm \pspm' ) \,   .   \lab{2.4}
\ee

Starting with the basic PB
\be
\{\pspm^i(\s),\pi_\pm^j ({\bar \s}) \}=\d^{ij} \d(\s-{\bar \s}) \, ,   \lab{2.5}
\ee
we can find that PB of the currents satisfies two independent
copies of KM algebras {\it without central charges}
\be
\{j_{\pm a},j_{\pm b} \} =f_{ab}{}^c j_{\pm c} \d \,   ,   \qquad
\{j_{+ a},j_{- b} \}=0 \,   .                                \lab{2.6}
\ee
We also have the relations
\be
\{\th_\pm ,j_{\pm a} \} =j_{\pm a} \d' \,  ,    \qquad
\{\th_\pm , j_{\mp a}\}=0 \,  ,                             \lab{2.7}
\ee
which imply
\be
\{{\cH}_c ,j_{\pm a} \} = \pm j_{\pm a} \d'  \,   .     \lab{2.8}
\ee

The total hamiltonian takes the form
\be
H_T = \int d\s [{\cH}_c +\sqrt{2} (A_+^a j_{-a}+ B_-^a j_{+a})] \,  . \lab{2.9}
\ee

The consistency conditions for the currents
\bea
&&{\dot j_{+ a}} =\{j_{+ a} ,H_T \}=  j_{+ a}' +\sqrt{2}
f_{ab}{}^c B_-^b j_{+ c} \,  ,  \nn\\
&&{\dot j_{- a}} =\{j_{- a} ,H_T \}= -j_{- a}' +\sqrt{2}
f_{ab}{}^c A_+^b j_{- c} \,   ,                \lab{2.10}
\eea
do not lead to new constraints, because the right hand sides of \eq{2.10}
are weakly equal to zero. In fact, the last equation means that
chiral currents are separately conserved
\bea
&&D_- j_{+ a} \equiv \pd_- j_{+ a} -f_{ab}{}^c B_-^b j_{+ c} =0 \,
,\nn\\
&&D_+ j_{- a} \equiv \pd_+ j_{- a} -f_{ab}{}^c A_+^b j_{- c} =0 \, ,  \lab{2.11}
\eea
or that both vector and axial vector currents are conserved.

The currents $j_{- a}$ and $j_{+ a}$  correspond to the arbitrary multipliers
$A_+^a$ and $B_-^a$ respectively in \eq{2.9}, and consequently they  are FCC.
Eqs. \eq{2.6} lead to the same conclusion.

Therefore, the classical theory has local non-abelian gauge symmetries, whose generators
$j_{\pm a}$ satisfies the corresponding PB algebra \eq{2.6}.

\subsection{Quantum theory}

In passing from the classical to the quantum domain,
we introduce the operators $\hat{\psi}_\pm^i$
instead of the field $\pspm^i$, replace the PB by the commutators
and define the composite operators using normal ordered prescription
\be
\hat j_{\pm a} =:\hat \pi_\pm t_a {\hat \psi}_\pm: \,  ,   \qquad
\hat \th_\pm=:\hat \pi_\pm {\hat \psi}_\pm' :  \,   .        \lab{2.12}
\ee
The gauge fields $A_+^a$ and $B_-^a$ will be considered as
classical background fields.

Then, instead of the PB algebra \eq{2.6} and \eq{2.7} we obtain corresponding
commutator algebra
\bea
&[\hat j_{\pm a} , \hat j_{\pm b} ] = i\hbar [f_{ab}{}^c \hat j_{\pm c} \d \pm
2\k \d_{ab} \d'] \,  , \qquad \, &[\hat j_{+ a} , \hat j_{- b} ] =0\,  ,   \nn \\
&[\hat \th_\pm ,\hat j_{\pm a}] = i\hbar \hat j_{\pm a} \d' \,  ,
     &[{\hat \th}_\pm , {\hat j}_{\mp a} ]=0 \,  ,        \lab{2.13}
\eea
with $\k={-\hbar \over 8\pi}$. For details of derivations see the Appendix.

In the quantum theory, as well as in the classical one, we also have a pair of
commuting KM algebras but this time with a {\it central charge}, in this case known
as Schwinger term. Therefore, the constraints $j_{\pm a}$ which were FCC
in the classical theory, become SCC operators ${\hat j}_\pm^a$ in the quantum
theory. This means that the theory is anomalous, because the classical
symmetry generated by FCC $j_{\pm a}$ is destroyed at the
quantum level. After quantization the theory obtains new degrees of freedom.

Note that under parity transformation $P: \, {\hat \psi}_\pm \rightarrow {\hat
\psi}_\mp$, so that $P \hat j_{\pm a} (\t, \s) P =\hat j_{\mp a} ( \t ,-\s)$ and
$P \hat \th_\pm ( \t,\s) P=-\hat \th_\mp (\t,-\s)$. Consequently, relations \eq{2.13}
with plus and minus indices are connected by parity
transformation. This means that our regularization scheme is left-right
symmetric, because the normal order prescription takes the
regularization role of the theory.

\section{GAUGE WZNW MODEL AS AN EFFECTIVE ACTION} 

The PB algebra \eq{2.6} is the  symmetry generator algebra,
because $j_{\pm a}$ are of the  FCC. The commutator algebra \eq{2.13} is the
algebra of dynamical variables (except the zero modes, see \cite{8}),
because there is a constant central term on the right hand side.
Our intention is to find the {\it effective theory} for these
variables, which means the quantum version of the action \eq{2.1}.

We introduce new  variables $J_{\pm a}$ and $\Th_\pm$ and
postulate that their classical PB algebra
\bea
&\{J_{\pm a} ,J_{\pm b} \}= f_{ab}{}^c J_{\pm c} \d \pm 2\k
\d_{ab} \d' \,   ,\qquad &\{J_{+a} ,J_{-b}\} =0 \,  ,\nn \\
&\{\Th_\pm ,J_{\pm a}\}= J_{\pm a} \d' \,  ,
&\{\Th_\pm ,J_{\mp a} \}=0 \,   ,          \lab{3.1}
\eea
is isomorphic to the
commutator algebra \eq{2.13} of the operators ${\hat j}_{\pm a}$
and ${\hat \th}_\pm$. We also define the canonical and the total
hamiltonian densities in analogy with  \eq{2.4} and \eq{2.9}
\be
{\cH}_c =\Th_+ - \Th_-  \,  , \qquad
{\cH}_T={\cH}_c +\sqrt{2} (A_+^a J_{-a} + B_-^a J_{+a})\,  .   \lab{3.2}
\ee

We should construct the canonical effective action $W$, for the
theory with PB algebra \eq{3.1} and with hamiltonian density \eq{3.2}.
In subsec.A we are going to find the expressions for the currents
and hamiltonian density in terms of the phase-space variables, and then
in subsec.B we will apply general canonical method \cite{6,7} to find the action $W$.

\subsection{Bosonic representation for the PB algebra} 
Let us "solve" eqs. \eq{3.1}, i.e. find the
expressions for the currents $J_{\pm a}$ and for the energy-momentum
tensor $\Th_\pm$ in terms of the coordinate $q^\a$ and the corresponding
momenta $\pi_\a$, which satisfies
\be
\{q^\a ,\pi_\b \} = \d^\a_\b \d \,  .                            \lab{3.3}
\ee
We will  start with the {\it ansatz}, that the currents are linear
in the momenta
\be
J_{\pm a} =- E_{\pm a}{}^\a ( \pi_{\a} +R_{\pm \a})  \,  ,    \lab{3.4}
\ee
where the coefficients $E_{\pm a}{}^\a$ and $R_{\pm \a}$ are the functions
of $q^\a$ only, and do not depend on the $\pi_\a$. We also suppose that the matrices
$E_{\pm a}{}^\a$ have inverses, which we denote by $E_{\pm \a}{}^a$.
The indices $\a,\b,...$ run over the same range as $a,b,..$.

Substituting \eq{3.4} into the first equation \eq{3.1}  we obtain an
equation linear in $\pi_\a$. The vanishing coefficient in front of momentum
gives
\be
E_{\pm b}{}^\b \pd_\b E_{\pm a}{}^\a - E_{\pm a}{}^\b \pd_\b E_{\pm b}{}^\a =
-f_{ab}{}^c E_{\pm c}{}^\a \,  ,                                \lab{3.5}
\ee
or equivalently
\be
\pd_\b E_{\pm \a}{}^c- \pd_\a E_{\pm \b}{}^c = f_{ab}{}^c E_{\pm \a}{}^a
E_{\pm \b}{}^b \,  .                                            \lab{3.6}
\ee
The second condition (term without $\pi$) yields
\be
E_{\pm a}{}^\a E_{\pm b}{}^\b [ \{ \pi_\a , R_{\pm \b} \} + \{R_{\pm \a} ,\pi_\b \} ] =
\pm 2\k \d_{ab} \d'  \,  .                                        \lab{3.7}
\ee
On the right side there is derivative of the delta function, so there must also be a derivative
on the left side, and we suppose that
\be
R_{\pm \a} = P_{\pm \a \b} (q) {q^\b}'  \,  .                      \lab{3.8}
\ee
Using this in \eq{3.7} we obtain two conditions,
\be
E_{\pm a}{}^\a E_{\pm b}{}^\b (P_{\pm \a \b} +P_{\pm \b \a}) =
\pm 2\k \d_{ab} \,  ,                                       \lab{3.9}
\ee
and
\be
E_{\pm a}{}^\a \pd_\g E_{\pm b}{}^\b P_{\pm \a \b} +
E_{\pm a}{}^\a \pd_\g (E_{\pm b}{}^\b P_{\pm \b \a}) +
E_{\pm a}{}^\a E_{\pm b}{}^\b (\pd_\b P_{\pm \a \g} -\pd_\a P_{\pm \b \g} )=0 \, ,   \lab{3.10}
\ee
because the coefficients in front of $\d'$ and $\d$ must vanishes separately.

If we define the symmetric tensor
\be
\g_{\pm \a \b} = E_{\pm \a}{}^a E_{\pm \b}{}^b \d_{ab} \,     \lab{3.11}
\ee
we can rewrite \eq{3.9} as $P_{\pm \a \b} +P_{\pm \b \a} =\pm 2\k \g_{\pm \a \b}$ and
find its general solution
\be
P_{\pm \a \b}= 2\k (\t_{\pm \a \b} \pm{1 \over 2} \g_{\pm \a \b} ) \,  ,     \lab{3.12}
\ee
where $\t_{\pm \a \b} =- \t_{\pm \b \a}$ is some antisymmetric tensor.
The first term is a solution of homogeneous part and the second one is
a particular solution of the full equation.

With the help of \eq{3.6} and \eq{3.12} we can obtain  from \eq{3.10}
\be
\pd_\a \t_{\pm \b \g}+\pd_\b \t_{\pm \g \a}+\pd_\g \t_{\pm \a \b} =
\mp {1\over 2} f_{abc} E_{\pm \a}{}^a E_{\pm \b}{}^b E_{\pm \g}{}^c  \, .   \lab{3.13}
\ee
Therefore, from the first eq. \eq{3.1} we got two relations,  \eq{3.6}
and \eq{3.13}. The first one is a condition on the $E_{\pm \a}{}^a$'s
and the second one defines  $\t_{\pm \a \b}$'s in terms of $E_{\pm \a}{}^a$'s.

From the second eq. \eq{3.1} we also obtain three equations
\be
E_{+a}{}^\a \pd_\a E_{-b}{}^\b - E_{-b}{}^\a \pd_\a E_{+a}{}^\b =0 \,  ,  \lab{3.14}
\ee
\bea
&&[-E_{+ a}{}^\a \pd_\a E_{- b}{}^\b P_{- \b \g} +
E_{-b}{}^\a \pd_\a E_{+a}{}^\b P_{+ \b \g} +
E_{+a}{}^\a \pd_\g E_{-b}{}^\b (P_{- \b \a}+P_{+ \a \b}) \, \nn\\
&&+E_{+ a}{}^\a E_{- b}{}^\b (-\pd_\a P_{- \b \g} +\pd_\b P_{+ \a \g}
+\pd_\g P_{- \b \a})] {q^{\g}}' =0 \,  ,       \lab{3.15}
\eea
and
\be
P_{+ \a \b}+ P_{- \b \a} =0 \,  ,              \lab{3.16}
\ee
as a coefficients in front of $\pi_\b \d$, $\d$ and $E_{+ a}{}^\a
E_{- b}{}^\b \d'$ respectively. From \eq{3.12}, \eq{3.16} and the symmetry
properties of $\t$ and
$\g$, follows
\be
\t_{+ \a \b} =\t_{- \a \b}, \qquad   \g_{+ \a \b} =\g_{- \a \b} \,  ,    \lab{3.17}
\ee
and consequently, from now we will  just call them $\t_{\a \b}$ and $\g_{\a \b}$, so
that \eq{3.12} becomes
\be
P_{\pm \a \b} =2\k (\t_{\a \b} \pm {1 \over 2} \g_{\a \b}) \,  .      \lab{3.18}
\ee
With the help of \eq{3.6} and \eq{3.11} we recognize $E_{\pm \a}{}^a$ as
vielbeins on the group manifold, and $\g_{\a \b}$ as the Cartan metric
in coordinate basis.

Eq. \eq{3.15} is a linear combination of equations \eq{3.13}, \eq{3.14} and
\eq{3.16}, and does not give anything new. We will discus eq. \eq{3.14} soon.

To make the geometric interpretation clearer we introduce a differential form notation.
Let us define the pair of Lie-algebra valued 1-forms
\be
v_\pm = t_a E_{\pm}{}^a{}_\a d q^\a  \,  ,                          \lab{3.19}
\ee
and the 2-form
\be
\t ={1 \over 2} \t_{\a \b} dq^\a dq^\b \,   .                       \lab{3.20}
\ee
Then equations \eq{3.6} become the Maurer-Cartan (MC) equations
\be
d v_\pm + v_\pm^2 =0 \,  .                             \lab{3.21}
\ee
They have a simple solutions in which the MC forms $v_\pm$ are
expressed in terms of group-valued fields $g_\pm$ as
\be
v_\pm =g_\pm{}^{-1} d g_\pm \,     .                     \lab{3.22}
\ee

In new notation equation \eq{3.13} with the help of \eq{3.17} obtains the form
\be
d \t = \mp {1 \over 3!} ( v_\pm , v_\pm^2 ) \,   ,          \lab{3.23}
\ee
where $(X,Y)$ is the Cartan inner product defined as $(t_a ,t_b)=\d_{ab}$,
so that in our normalization $(t_a ,t_b)=-2 tr\{t_a t_b \}$.

From the second eq.\eq{3.17} and \eq{3.11} we have
\be
(v_+ , v_+  ) = (v_- , v_-) \,  ,                      \lab{3.24}
\ee
and from \eq{3.23}
\be
(v_+ ,v_+^2) =-(v_- ,v_-^2) \,  .                    \lab{3.25}
\ee
The Cartan-Killing form $(X ,Y)$ is invariant under the adjoint
action of the group element $g\,$: $(g^{-1} X g , g^{-1} Y g) =(X,Y)$,
so we can conclude that $v_+$ and $v_-$ are  connected as
\be
v_- =- g^{-1} v_+ g \,   ,                          \lab{3.26}
\ee
where $g$ is a group valued field. Substituting \eq{3.22} in \eq{3.26}
we easily obtain that $g=g_- =g_+^{-1}$ and finally
\be
v_+=gdg^{-1}\, , \qquad   v_-= g^{-1} d g \,    .                \lab{3.27}
\ee

Let us now come back to equation \eq{3.14}. After some transformations we can write
it as
\be
dv_- = g^{-1} d v_+ g  \,  ,                                \lab{3.28}
\ee
or with the help of \eq{3.21} as $v_-^2 = g^{-1} v_+^2 g$. Therefore, it is not a new
relation but the consequence of the  \eq{3.26}.

The final result for the current components is
\be
J_{\pm a} =- E_{\pm a}{}^\a [\pi_\a + 2\k (\t_{\a \b} \pm {1 \over 2}
\g_{\a \b}) {q^\b}'] \,  .                                   \lab{3.29}
\ee

We still need to "solve" the last two equations \eq{3.1}. Because all
expressions with the opposite chirality commute, we will take
$\Th_\pm =\Th_\pm (J_{\pm a})$. The group invariant expressions
\be
\Th_\pm =\pm {1 \over 4 \k} \g^{\a \b} J_{\pm a} J_{\pm b} \,   ,        \lab{3.30}
\ee
representing the components of energy-momentum tensor, is the solution
we are looking for. It is enough to use only the first two relations \eq{3.1} and
not the expressions \eq{3.29}, to check that \eq{3.30} satisfies
the last two equations \eq{3.1}.

\subsection{Effective action }

We are ready to construct the effective action based on the general canonical formalism.
It takes the standard form
\be
W(q,\pi ,A,B) =\int d^2 \xi ( \pi_\a {\dot q}^\a -{\cH}_T) \,   ,            \lab{3.31}
\ee
where the total hamiltonian  ${\cH}_T=\Th_+ -\Th_-  +\sqrt{2} (A_+^a J_{-a} + B_-^a
J_{+a})$, is defined in \eq{3.2}. The expressions for the current
components and for the energy-momentum tensor components in terms
of canonical pairs $(q^\a , \pi_\a)$ are defined in \eq{3.29} and \eq{3.30}.

To find the usual second-order form of the action, we will eliminate
the momentum variables $\pi_\a$ on their equations of motion
\be
{\dot q}^\a - {1 \over  2\k} \g^{\a \b} (J_{+ \b} + J_{- \b} ) +
\sqrt{2} (A_+^\a + B_-^\a )=0 \,  ,                    \lab{3.32}
\ee
where $A_+^\a = E_+{}^\a{}_a A_+^a $, $B_-^\a = E_-{}^\a{}_a
B_-^a$  and $J_{\pm \a} = -E_{\pm \a}{}^a J_{\pm a}$ . With the help of
\eq{3.29} we have on the equations of motion
\be
J_\pm^\a = \sqrt{2} \k (\pd_\pm q^\a + A_+^\a + B_-^\a ) \,   .  \lab{3.33}
\ee
Substituting this in \eq{3.31}, after some calculations we find
\be
W(q,A,B) =W_0 (q)+W_1 (q,A,B) \,  ,              \lab{3.34}
\ee
where
\bea
W_0 (q)=&&-\int d^2 \xi P_{- \a \b} \pd_- q^\a \pd_+ q^\b =
\k \int d^2 \xi (\g_{\a \b} -2 \t_{\a \b} ) \pd_- q^\a \pd_+ q^\b
\,  ,\nn \\
W_1 (q,A,B) =&&2\k \int d^2 \xi [\pd_- q^\a E_{+ \a}{}^a A_{+a} +
\pd_+ q^\a E_{- \a}{}^a B_{- a} + A_+^a E_{+ a}{}^\a E_{- \a}{}^b B_{- b} \nn\\
&&+\fr{1}{2}(A_+^a A_{+ a} +B_-^a B_{- a})]  \,  .               \lab{3.35}
\eea

It is possible to add to the effective action  some local
functional, depending on the fields $A_+$ and $B_-$.
In order to cancel the last term in \eq{3.35}  we add
\be
\D W =-\k \int d^2 \xi [A_+^a A_{+a} + B_-^a B_{-a} ]\,   ,      \lab{3.36}
\ee
and get
\be
W_{l,r} =W + \D W \,   ,                           \lab{3.37}
\ee
where the meaning of the indices $l,r$ will be clear in Sec. V.

In the differential form notation, with Cartan inner product normalization,
using the Stoke's theorem we have
\bea
W_0 (v)=&&\fr{1}{2}\k\int_\S ({}^*v_\pm,v_\pm)\pm
\fr{1}{3}\k\int_M (v_\pm,v_\pm^2) \,   ,             \nn\\
(W_1 +\D W)(v,A,B,)=&&\k \int_\S [(v_+, A- {}^*A) -
(v_-,B+{}^*B) \nn\\
&&-\fr{1}{2} (B+{}^*B, g^{-1} (A-{}^*A) g) ]\, .            \lab{3.38}
\eea
Here, $W_0(v)$ is the well known WZNW model \cite{2,9}.
As defined in \eq{3.19}, $v_\pm$ are the Lie algebra valued 1-forms
and ${}^*v_\pm$ are the dual of $v_\pm$. Both expressions for $W_0(v)$ are
equal on the basis of \eq{3.24} and \eq{3.25}. The first term is the action
of the non--linear $\s$--model, while the second one is the topological
Wess--Zumino term, defined over a three--manifold $M$ whose boundary is the
spacetime: $\pd M=\S$. For our value of $\k= -{\hbar \over 8 \pi}$ it reads
\be
{\hbar \over 24 \pi} \int_M (g^{-1} dg , g^{-1} dg g^{-1} dg )\,  ,
                                \lab{3.39}
\ee
so that it is well-defined modulo $2 \pi \hbar n$, where $n \in Z$ is a
winding number. We want to stress, that the proper value of
$\k$ for which quantum theory is single-valued is determined from the
central charge,  which we obtained using just the normal ordered prescription
in quantum fermionic theory.

The expression for $W_1+ \D W$ is the regular part of the gauge extension of the WZNW action.
It has been obtained in the process of consistent gauge invariant extension of
the WZNW model \cite{10}.

Starting with the PB algebra \eq{3.1} and eq. \eq{3.2}, we obtained the action
\eq{3.38} for any value of the constant $\k$. But, the uniqueness of the
irreducible representation of the KM algebra, as well as topological arguments
\cite{2} force it just to the same value as in our normal-order approach.

\section{Bosonization} 

\subsection{ The chiral currents bosonization rules}

In the previous section we obtained the effective action $W$
for massless fermi theory in the external gauge fields. This is
equivalent, to solving the functional integral
\be
Z_F (A,B) = \int d \psp d \psm e^{i S(\psp, \psm, A, B)} \,  .  \lab{4.1}
\ee
The final result depends only on the background fields $A$ and $B$. The
bosonic expression for $W$ depends not only on $A$ and $B$, but
also on some auxiliary  fields, $q^\a$ and $\pi_\a$ in the hamiltonian and $g$
in the lagrangian case. So, after integration over auxiliary
fields we can eliminate them and obtain
\be
Z_B (A,B) = \int d\pi dq e^{i W(q,\pi,A,B)} \, , \qquad
Z_B (A,B) = \int dg e^{i W_{l,r}(v,A,B)} \,  ,          \lab{4.2}
\ee
for the hamiltonian and the lagrangian approach respectively.
We are not going to do the integrations explicitly, because they lead to
non local expression in terms of $A$ and $B$, even in the abelian case.
We only conclude that $Z_F$ should  be proportional to $Z_B$
\be
Z_F \sim Z_B \,  .   \lab{4.3}
\ee
This functional integral identity admits interpretation in
terms of bosonization. Differentiating \eq{4.3} with respect to
$B_-^a$ and $A_+^a$ and setting $A_+ =0=B_-$, we obtain the bosonization
rules for chiral currents in non-abelian theory.

If we choose the expression $W(q,\pi,A,B)$ from \eq{3.31} as an
effective action we get the hamiltonian bosonization rules
\be
i\psi_\pm^* t_a \pspm \to -E_{\pm a}{}^\a (\pi_\a + P_{\pm \a \b}
{q^\b}') \,  ,              \lab{4.4}
\ee
and if we choose $W_{l,r} (v,A,B)$ from  \eq{3.38}, we
get the well known lagrangian bosonization rules
\be
i\psp^* t_a \psp \to - \sqrt{2} \k g \pd_+ g^{-1} \,  \quad
i\psm^* t_a \psm \to - \sqrt{2} \k g^{-1} \pd_- g \,  .  \lab{4.5}
\ee

\subsection{The chiral densities bosonization rules}

It is possible to add the mass term
\be
\bpsi \psi = \psp^* \psm + \psm^* \psp \,   \lab{4.6}
\ee
to the action \eq{2.1} and find the corresponding expression in terms
of the bosonic variables. Let us introduce the chiral densities
\be
\r_\pm  =\bpsi {1 \pm \g_5 \over 2} \psi = \pspm^* \psi_\mp
=-i\pi_\pm \psi_\mp\,  .                  \lab{4.7}
\ee
All expressions are matrices with indices $i,j$ (for example $\r_\pm^{ij} =
-i \pi_\pm^i \psi_\mp^j$) but we will omit them for simplicity.

It is easy to find the PB between the chiral densities and the
currents $j_{\pm a}$, defined in \eq{2.3}
\be
\{j_{\pm a} ,\r_\pm \} =(\r_\pm t_a) \d \, ,  \quad
\{j_{\pm a} , \r_\mp \} = -(t_a \r_\mp ) \d \,  .     \lab{4.8}
\ee

In the quantum theory the central term does not appear, so the
commutation relations, up to $i \hbar$, are the same as PB \eq{4.8}.
The commutators with the currents ${\hat j}_{\pm a}$  completely
define the expressions for ${\hat \r}_\pm$.

For the bosonic representations, instead of ${\hat \r}_\pm$ we
introduce the corresponding matrix valued expressions $\Upsilon_\pm$,
depending on the bosonic variables. Their PB algebra with
$J_{\pm a}$ should be isomorphic to the operators algebra of
${\hat \r}_\pm$ with ${\hat j}_{\pm a}$
\be
\{J_{\pm a} , \Upsilon_\pm \} =(\Upsilon_\pm t_a) \d \,   ,  \quad
\{J_{\pm a} , \Upsilon_\mp \} = - (t_a \Upsilon_\mp ) \d  \,  .       \lab{4.9}
\ee
Both relations \eq{4.9} will give the same result, so we will
solve only the first one.

Assuming that $\Upsilon_\pm$ does not depend on the momenta, and
using the expression \eq{3.29} for the current, we obtain the equation
\be
E_{\pm a}{}^\a \pd_\a \Upsilon_\pm = \Upsilon_\pm t_a \,   .       \lab{4.10}
\ee
Multiplying it with $E_\pm{}^a{}_\b dq^\b$,  with the help of
\eq{3.19} we get
\be
\Upsilon_\pm^{-1} d \Upsilon_\pm = v_\pm \,  .               \lab{4.11}
\ee
Comparing last equation with \eq{3.27}, we conclude that
\be
\Upsilon_+ = M g^{-1} \, , \qquad   \Upsilon_- = M g \,  ,        \lab{4.12}
\ee
where $M$ is a constant. Because $g$ is dimensionless, $M$ has
dimension of mass. At the end, we can complete our bosonization
formulae for the chiral densities
\be
\bpsi \psi \to \Upsilon_+ + \Upsilon_- =M (g + g^{-1}) \,  , \quad
\bpsi \g_5 \psi \to \Upsilon_+ + \Upsilon_- = M (g^{-1} - g) \, .    \lab{4.13}
\ee
They are the same for the hamiltonian and lagrangian case, because
they do not depend on the momenta and agree with those of reference \cite{2}.

\section{Canonical approach to Anomalies}  

\subsection{From Schwinger term to anomalies}

The Schwinger term breaks the symmetries, changing the
generators from the FCC to the SCC. It is interesting to investigate its
influence on the transformation properties of the effective action.

Let us extend the previously described general canonical
method for constructing the effective action, from the known PB algebra
to the case with the central term. The basic idea of that
approach was, that the action
\be
W = \int d^2 \xi [\pi_\a {\dot q}^\a -{\cH}_c - \l^m G_m ] \,  \lab{5.1}
\ee
is invariant under gauge transformations
\be
\d F(\s) =\{ F(\s) , \int d{\bar \s} \ve ({\bar \s}) G_m ({\bar \s}) \} \,  \lab{5.2}
\ee
of any quantity $F(\pi, q)$, and
\be
\d \l^m  ={\dot \ve}^m -  {\ve^n}'h_n^m -  \ve^n \l^k f_{kn}{}^m \,        \lab{5.3}
\ee
of lagrange multipliers $\l^m$, if $G_m$ are the FCC
\be
\{ G_m , G_n \} = f_{mn}{}^p G_p \d \,  , \qquad
\{ {\cH}_c , G_m \} = h_m{}^n G_n \d'   \,  .    \lab{5.4}
\ee
Here we adopted the notation appropriate for field theory.

In the case where the central term is present the first equation \eq{5.4} becomes
\be
\{ G_m , G_n \} = f_{mn}{}^p G_p \d + \D_{mn} (\s, {\bar \s}) \,  . \lab{5.5}
\ee
We want to preserve the gauge transformations of the fields \eq{5.2} and
\eq{5.3}. Then the Schwinger term appears in the variation of the effective
action
\be
\d W =- \int d\t d\s \int d {\bar \s} \l^m (\s) \D_{mn} (\s, {\bar \s})
\ve^n({\bar \s}) \,  .                             \lab{5.6}
\ee

The method, we have used, works only for actions linear in the gauge fields
$\l^m$. We can add arbitrary local functional $\D W(\l^m)$ to the effective
action and obtain the general expression for the anomaly
\be
{\cal A}_n (\s,\t) ={\d W \over \d \ve^n (\s, \t)} = \int d {\bar \s}
\D_{nm} (\s,{\bar \s}) \l^m ({\bar \s}, \t)
+{\d \over \d \ve^n(\s, \t)} \D W  \,  .       \lab{5.7}
\ee
The non trivial part of the anomaly is  proportional to the Schwinger term,
because it breaks the symmetry and measures non-invariance of the effective action.

We want to emphasize, that we do not need the expression for the generators $G_m$ in
terms of $\pi_\a$ and $q^\a$ to obtain \eq{5.6} and \eq{5.7}. It is enough to know only the
central term, and the anomaly will depend on the gauge field
$\l^m$, but not on the phase space coordinate.

\subsection{The left-right and the axial anomalies}

We can apply this method to the classical fermionic theory with
$$
\ba{lllll}
G_m=   & j_{+a},   & j_{-a} ,    \\
\l^m=   & \sqrt{2} B_-^a ,   & \sqrt{2} A_+^a ,    \\
\ve^m= & \b^a , & \a^a ,
\ea
$$
and obtain the well known transformations under the local gauge group
$G_l \times G_r$
\bea
&&\d \psm =- \a \psm \,  ,  \qquad  \d A_+ =\pd_+ \a -[A_+ , \a] \,  , \nn\\
&&\d \psp =- \b \psp \,  ,  \qquad  \d B_- =\pd_- \b -[B_- , \b] \,   \lab{5.8}
\eea
where $\a=\a^a t_a$ and $\b=\b^a t_a$.

In the case of the bosonic theory we have
$$
\ba{lllll}
G_m=   & J_{+a},   & J_{-a} ,    \\
\l^m=   & \sqrt{2} B_-^a ,   & \sqrt{2} A_+^a ,    \\
\ve^m= & \b^a , & \a^a ,
\ea
$$
and
\be
\D_{mn} (\s, {\bar \s}) \to \pm 2\k \d_{ab} \d' (\s - {\bar \s})\,    \qquad
(m \to (a, \pm) \,  , n \to (b, \pm))  \,  .             \lab{5.9}
\ee
The local gauge transformations for the fields $A_+$ and $B_-$ are the
same as in \eq{5.8} and for matter fields we have
\be
\d q^\a = -\b^a E_{+a}{}^\a -\a^a E_{-a}{}^\a \,   , \lab{5.10}
\ee
which yields
\be
\d g = \b g - g \a \,                        \lab{5.11}
\ee
(see \cite{8} and the second ref. \cite{7}).

Under these transformations we have from \eq{5.6} and \eq{5.9}
\be
\d W =- 4 \sqrt{2} \k \int d^2 \xi tr\{ \b B_-' - \a A_+'\}\,  .     \lab{5.12}
\ee
When we include the transformation of the $\D W$ \eq{3.36}
\be
\d (\D W)= -4\k \int d^2 \xi tr\{\a \pd_+ A_+ + \b \pd_- B_- \} \,       \lab{5.13}
\ee
we obtain
\be
\d W_{l,r}= -4\k \int d^2 \xi tr\{\a \pd_- A_+ + \b \pd_+ B_- \} \,  ,
\qquad {\cal A}_l=2\k \pd_- A_+ \,  ,
 \qquad {\cal A}_r=2\k \pd_+ B_+ \,   ,               \lab{5.14}
\ee
where $W_{l,r} =W+ \D W$ as in \eq{3.37}. In this case both left and right
symmetries are anomalous, which we denoted by indices $l,r$.

We can add the finite local counterterm
\be
\D W_{ax} =-4\k \int d^2 \xi tr\{ A_+ B_-\} \,   ,             \lab{5.15}
\ee
and shift anomaly from left-right to axial one. The redefined effective
action $W_{ax} = W_{l,r} + \D W_{ax}$ is invariant under the vector gauge
transformation: $\a=\b=\ve_v$
\be
\d_v W_{ax} =0 \,    ,                                        \lab{5.16}
\ee
but not under the axial one: $\a=-\b=\ve_{ax}$
\be
\d_{ax} W_{ax} =-8\k \int d^2 \xi tr\{\ve_{ax} {}^*F\} \,   ,
\qquad {\cal A}_{ax}=4\k {}^*F \,  ,              \lab{5.17}
\ee
where
\be
{}^*F = \fr{1}{2} \ve^{\m \n} F_{\m \n} \,    ,           \lab{5.18}
\ee
and
\be
F_{-+} = \pd_- A_+ - \pd_+ B_- + [A_+ , B_- ] \,  .         \lab{5.19}
\ee
The second equation  \eq{5.17} is the well known result for axial anomaly.

The non-invariance of the effective action is a consequence of the
non-conservation of the currents. The hamiltonian equations of
motion have anomalous divergent currents, instead of the conserved one
in \eq{2.11}. Taking into account $\D W$ and $\D W_{ax}$, the same expressions for
${\cal A}_l\,$, ${\cal A}_r\,$ \eq{5.14} and for ${\cal A}_{ax}\,$
\eq{5.17} can be obtained.

\section{Concluding remarks} 

We presented here a complete and independent derivation of the two
dimensional gauged WZNW model, using the hamiltonian methods. We also obtained
hamiltonian and lagrangian  non-abelian bosonization rules and
the expression for the anomalies.

We started with canonical analysis of the theory of massless
chiral fermions coupled to the external gauge field. We found
that there are FCC $j_{\pm a}$, whose  PB satisfies two independent
copies of KM algebras without central charges. In passing to the
quantum theory, the central term appears in the commutation relations
of the operators ${\hat j}_{\pm a}$, which changes the nature of constraints:
they become SCC instead of FCC.

We define the new effective theory, postulating the PB of the constraints
and hamiltonian density. Particularly, we require that PB algebra of the
classical bosonic theory should be isomorphic to the commutator algebra of
the quantum fermionic theory. Then we found the
representation for the currents and hamiltonian density in terms of
phase space coordinates. Finally, we derived effective action using
general canonical formalism and obtained the gauged WZNW model.
We want to stress, that we also got the topological Wess-Zumino term.
The tensor $\t_{\pm \a \b}$, as its origin, appears in our approach as a
general solution of the homogeneous part of the equation \eq{3.9}.
The coefficient in front of the Wess-Zumino term is defined by the numerical
value of the central charge and gives a correct expression for the winding
number.

Once we establish the connection between the fermionic and the bosonic
theories, it was easy to find the bosonization rules, just
differentiating generating functionals with respect to the background fields.
Beside usual bosonization rules,
we also got the hamiltonian ones, expressing the currents $J_{\pm a}$
in terms of both coordinate $q^\a$ and momentum $\pi_\a$. After
elimination of momenta on the equations of motion, we came back to
the conventional bosonization rules.

The algebra of the currents $J_{\pm a}$ is the
basic PB algebra. Knowing its representation in terms of $q^\a$ and $\pi_\a$
we can find the representation for all other quantities from their
PB algebra with the currents. As an example we found the bosonization rules
for the chiral densities.

The canonical approach is very suitable for calculation of the anomaly.
The general formula \eq{5.7} expresses the anomaly as a function of
the Schwinger term. The normal ordering prescription for the quantum
operators takes the role of left-right symmetric regularization
scheme. So, we obtain both left and right anomalies.
By adding the finite local counterterm we in fact changed the regularization
scheme, and shifted the anomaly from the left-right symmetric to the axial one.

The Schwinger term, and consequently the WZNW model
and the anomaly  have the correct dependence on
Planck's constant $\hbar $, because $\k$ is proportional to $\hbar$. The
fact that $\hbar$ arise in the classical effective theory, shows its
quantum origin.

The canonical approach of this paper can be applied to the other symmetries. If we
take the diffeomorphism transformations instead of the non-abelian ones,
then we will get Virasoro algebra, 2D induced gravity and conformal
anomaly instead of KM algebra, WZNW model and axial anomaly. The work of
this program is in progress and will be publish separately.

\section*{Acknowledgment} 

This work was supported in part by the Serbian Science Foundation,
Yugoslavia.

\appendix 

\section{Normal ordering and Schwinger term} 

In this Appendix we will derive the expression for the central
terms in the commutation relations \eq{2.13}.

We define currents ${\hat j}_{\pm a}$ as a quantum
operators and introduce normal ordering prescription.
Usually, it is convenient to employ the Fourier expansion
of the fields, identifying  the modes as a creation and
annihilation operators with respect to Fock vacuum state.

Following \cite{11} we prefer to decompose operators in positive
and negative frequencies in the position space. We introduce
two parts of delta function
\be
\d^{(\pm)} (\s) = \int_{- \infty}^{\infty} {d k \over 2 \pi}
\th(\mp  k) e^{ik (\s \mp i \ve)} = {\pm i \over 2\pi (\s \pm i
\ve)}  \,  , \quad (\ve > 0)         \lab{A1}
\ee
so that $\d (\s) = \d^{(+)}(\s) +\d^{(-)}(\s)$ with the
following properties
\be
\d^{(\pm)}(-\s) = \d^{(\mp)}(\s) \, , \qquad
(\d^{(+)})^2 - (\d^{(-)})^2 ={-i \over 2\pi} \d' \,  .  \lab{A2}
\ee

Then for any operator ${\hat \O}(\t,\s)$ we can perform the
splitting
\be
{\hat \O}^{(\pm)} (\t,\s) =\int_{-\infty}^\infty \d {\bar \s}
\d^{(\pm)} (\s -{\bar \s}) {\hat \O}(\t,{\bar \s}) \,  ,  \lab{A3}
\ee
where ${\hat \O} ={\hat \O}^{(+)} + {\hat \O}^{(-)}$.

Now, we adopt ${\hat \pi}_+^{(-)}$ and ${\hat \psi}_+^{(-)}$ as creation
operators and ${\hat \pi}_+^{(+)}$ and ${\hat \psi}_+^{(+)}$ as
annihilation operators
\be
{\hat \pi}_+^{(+)}|0\rangle = {\hat \psi}_+^{(+)} |0 \rangle =0 \,
 , \quad   \langle 0|{\hat \pi}_+^{(-)} = \langle 0|{\hat
\psi}_+^{(-)}=0 \,  .            \lab{A4}
\ee
To preserve symmetry under parity transformations,  we define
creation and annihilation operators for ${\hat \pi}_-$ and ${\hat
\psi}_-$ in an opposite  way (with index $(+)$ for a
creation and index $(-)$ for an annihilation operator).
Then in the both cases, the normal order for products of operators
means that annihilation operators are placed to the right of the
creations one.

From the basic commutation relations $[{\hat \psi}_\pm^i ,{\hat \pi}_\pm^j ]
=i\hbar\d^{ij} \d$, we can conclude that the only non trivial
parts are
\be
[ {\hat \psi}_+^{(\pm)} (\s) , {\hat \pi}_+^{(\mp)}({\bar \s})]
=i\hbar\d^{(\pm)} (\s - {\bar \s}) \,  , \quad
[ {\hat \psi}_-^{(\pm)} (\s) , {\hat \pi}_-^{(\mp)}({\bar \s})]
=i\hbar\d^{(\pm)} (\s - {\bar \s}) \,  .     \lab{A5}
\ee
After some calculation it is possible to check the commutator
algebra \eq{2.13}. Because the central term is the only possible difference
compared to the PB algebra,  the easiest way to confirm \eq{2.13} is to start
with expression
\be
[{\hat j}_{\pm a} , {\hat j}_{\pm b} ]= i\hbar (f_{ab}{}^c
{\hat j}_{\pm c} \pm \D_{ab} ) \,  ,     \lab{A6}
\ee
and find its vacuum expectation value. With the help of
\eq{A2} and with the convention $tr\{t_a t_b\}= -\fr{1}{2} \d_{ab}$,
we obtain
\be
\D_{ab} = {\hbar \over 2 \pi} tr \{t_a t_b \} \d' = 2 \k \d_{ab}
\d' \, ,   \qquad (\k ={-\hbar \over 8 \pi} )    \lab{A7}
\ee
proving the first relation \eq{2.13}.

Commutators $[{\hat j}_+, {\hat j}_-]$ and $[{\hat \th} , {\hat j}]$ do not have
central extensions.


\end{document}